\documentclass[12,leqno]{article} 

\usepackage{graphics}

\def\m#1{$#1$}

\def\sgn{\;{\rm sgn}\;}
\newcommand{\beq}{\begin{equation}}
\newcommand{\eeq}{\end{equation}}

\newcommand{\beqs}{\begin{eqnarray}}
\newcommand{\eeqs}{\end{eqnarray}}

\newcommand{\DOE}{This work was supported in part by U.S.Department 
of Energy grant No. DE--FG02-91ER40685}

\newcommand{\half}{\frac{1}{2}}
\newcommand{\eps}{\epsilon}

\def\hattilde#1{\hat{\tilde #1}}

\begin{document}
\input{epsf}

\centerline{\bf\large Parton Model from Bi-local Solitonic Picture of
the Baryon in two-dimensions
}

\centerline{V. John\footnote{vjohn@pas.rochester.edu}, 
G. S. Krishnaswami\footnote{govind@pas.rochester.edu} and 
S. G. Rajeev\footnote{rajeev@pas.rochester.edu}}

\begin{center}
   {\it Department of Physics and Astronomy, University of Rochester, 
    Rochester, New York 14627} \\
   \vspace{.5cm}

   \vspace{2cm}
   {\large\bf Abstract}
\end{center}

We study a previously introduced bi-local gauge invariant 
reformulation of two dimensional QCD, called 2d HadronDynamics. The 
baryon arises as a topological soliton in HadronDynamics. 
We derive an interacting parton model from the soliton model,
thus reconciling these two seemingly different points of view.
The valence quark model is obtained as a variational approximation
to HadronDynamics. A succession of better approximations to the soliton
picture are obtained. The next simplest case corresponds to a system of 
interacting valence, `sea' and anti-quarks. We also obtain this `embellished' 
parton model directly from the valence quark system through a unitary
transformation. Using the solitonic point of view, we estimate the
quark and anti-quark distributions of 2d QCD. Possible applications 
to Deep Inelastic Structure Functions are pointed out.


\vspace{.5 cm}

\pagebreak


\section{Introduction}

There are currently two distinct points of view on what a baryon is.
One may be traced back to the quark model. In the other point of
view, baryons arise as solitons of low energy effective local field theories
of mesons, an idea that may be traced back to the Skyrme Model \cite{skyrme}.
In previous work by one of us, it was shown that in two dimensions,
there is an exact description of QCD as a {\it bi-local} theory of
mesons \cite{2dqhd}. This description, called 2 dimensional HadronDynamics,
is not a low energy effective theory, but is equivalent to 
2d QCD for all energies and numbers of colors. The exact description of
the baryon is as a topological soliton of this bi-local theory.
At the other extreme, we studied a 2 dimensional 
interacting quark model for the structure of the baryon \cite{ipm}.
The question then is whether the quark model picture can be
derived from the exact bilocal solitonic picture in two dimensions.
Here we derive the interacting valence quark model 
as a variational approximation to this bi-local soliton theory.
Moreover, we find a succession of increasingly accurate variational
approximations to the soliton model. The next simplest case turns out be a 
system of interacting valence, `sea' and anti-quarks. 
We show that this `embellished' parton model could also have been obtained 
directly via a unitary transformation applied to the valence quark model,
recovering the `Bogoliubov' transformation introduced in \cite{ipm}.
Thus we have a reconciliation of the exact bi-local soliton model with 
the simpler relativistic parton picture of the baryon, in 2 dimensions.

The main advantage of this new point of view is that the
semi-classical approximation of HadronDynamics corresponds to the large 
\m{N_c} limit of 2d QCD, and so is capable of describing non-perturbative
phenomena such as the struture of hadrons. Thus this reconciliation
between the soliton and parton pictures is more than just a mathematical
correspondence. We illustrate its usefulness by calculating approximately, 
the quark and anti-quark distributions of the 
baryon in 2d QCD. In fact, due to the correspondence we establish here,
the approximate solutions of the quark models presented in \cite{ipm}
are actually also approximate solutions of 2d QCD. Thus, in this paper, we
will focus more on the passage from the bi-local soliton theory to
quark models, rather than their actual solution. As mentioned in \cite{ipm},
our results agree well with the direct numerical solutions
of Hornbostel et. al. \cite{hornbostel}. To summarize,
we find that the valence quark 
approximation is accurate not only in the non-relativistic limit, but also
in the ultra-relativistic chiral limit. In particular, we find that in the
chiral limit, the valence quark approximation is exact, for $N_c \to \infty$.

As an aside, we speculate on the possible phenomenological implications of
the above model. In Deep Inelastic Scattering \cite{dis}, the transverse 
momenta of the partons is small compared to their longitudinal momenta. 
Moreover, the observables of interest, the baryon structure functions, depend 
only on the proton momentum $P$ and photon momentum $q$, which lie in a 
two-dimensional time like hypersurface, spanned by time and the beam
direction. Thus, there must be an effective 2 dimensional theory
that describes the structure of the proton as measured in Deep Inelastic 
Scattering. Two dimensional HadronDynamics has the correct 
symmetries to be a candidate for an approximate description of the
`relevant' interactions of the quarks, in such an effective action. 
This model can be thought of as a representative example of such an 
approximate effective action. The variational and many-body techniques 
developed here should be useful in understanding any such 2 
dimensional effective action.


\section{Two Dimensional Quantum Hadron Dynamics}

Let us begin with a summary of Two dimensional Quantum HadronDynamics 
\cite{2dqhd}. 
2d QCD is quantized in the null gauge $A_- = 0$ in null coordinates. The 
elimination of longitudinal gluons leads to a linear potential between 
quark fields $a, a^{\dag}$ which satisfy canonical anti-commutation relations.
The resulting hamiltonian is:

\vspace{.1in}

\centerline{ \m{
{H \over N_c} = \int dx a^{\dag a  i}  \half[\hat p+{m^2\over \hat p}]   
a_{a i} - {g^2 \over 2 N_c} \int \half |x-y|
:a^{\dag a i}(x) a_{ a j}(x):
:a^{\dag b j}(y) a_{ b i}(y):dxdy.
}}

\vspace{.1in}

\noindent `g' is a coupling constant with the dimensions of mass,
$i, j$ are color indices and $a,b$ are flavour indices. Define 
the bilocal gauge invariant variable 
\m{\hat M^a_b(x,y)={2\over N_c}:a_{b i}(x) a^{\dag a i}(y):}.
The points \m{x,y} are null separated. The operators \m{\hat M^a_b(x,y)} 
form a complete set of observables in the color singlet sector of two 
dimensional QCD. They provide a (projective) unitary irreducible 
representation of the infinite dimensional unitary Lie Algebra:

\beqs
[\tilde {\hat M}^a_b(p,q),\tilde {\hat M}^c_d(r,s)] =
{1\over N_c} ( \delta_b^c 2\pi \delta(q-r)[\delta^a_d\sgn(p-s)
+ \tilde {\hat M}^a_d(p,s)]) \cr
\noindent - {1 \over N_c} (\delta_d^a 2\pi \delta(s-p)
[\delta^c_b\sgn(r-q)+\tilde {\hat M}^c_b(r,q)] ).
\eeqs

Note that the commutators are of order \m{1\over N_c} so that the large
 \m{N_c} limit is a sort of classical limit: \m{1\over N_c} plays the
 role that \m{\hbar} does in an ordinary field theory. In this 
classical limit the above commutators are replaced by the Poisson Brackets 
of a set of classical dynamical variables \m{M^a_b(x,y)}.
It was shown that the phase space of this system is 
an orbit of the unitary group, an infinite dimensional Grassmannian 
\m{Gr_1}. It is the set of all hermitean operators \m{M} with integral kernel 
\m{M^a_b(x,y)} satisfying the quadratic constraint \m{[\eps + M]^2~=~1},
with \m{\int |M(x,y)|^2 dxdy < \infty}. \m{Gr_1} is a curved 
manifold with connected components labelled by an integer.
 The quadratic constraint is just the Pauli principle for 
quarks: the density matrix \m{(\rho = \half (1 - M - \eps))} 
must be a projection operator. Here $\eps$ is the Hilbert transform operator,
diagonal in momentum space, with $\tilde\eps(p,p) = \sgn(p)$.

So in the large \m{N_c} limit, our problem reduces to solving the
equations of motion obtained from the hamiltonian and Poisson Brackets:

\beq
{E[M] \over N_c} = -{1\over 4}\int [p+{{\mu_a}^2\over p}]\tilde 
M^a_a(p,p){dp\over 2\pi} +{\tilde g^2\over 8}\int 
M^a_b(x,y)M^b_a(y,x)|x-y|dxdy
\eeq

\beqs
{1\over 2i}\{M^a_b(x,y),M^c_d(z,u)\}= 
\delta_b^c\delta(y-z)[\eps^a_d(x,u)+M^a_d(x,u)] \cr
\noindent - \delta_d^a\delta(x-u)[\eps^c_b(z,y)+M^c_b(z,y)].
\eeqs

\noindent The parameter \m{\mu_a^2} is related to the current 
quark masses \m{m_a} by a finite
renormalization: \m{\mu_a^2=m_a^2-{\tilde g^2\over \pi}} and
\m{\tilde{g}^2 = g^2 N_c}. Though the hamiltonian is quadratic, this is a 
non-linear interacting theory since the phase space is a curved 
manifold due to the constraints on \m{M(x,y)}. The linearization of the 
equation of motion around the vacuum \m{M^a_b = 0} describes an infinite 
number of free mesons, and 't Hooft's integral equation \cite{thooft} 
for the meson masses was recovered.

What kind of solution to this theory  represents the baryon? The quantity 
\m{
	B=-\half\int M^a_a(x,x)dx
}
was shown to be an integer, a topological invariant and hence 
conserved under time evolution. We see that $B$ is in fact
baryon number. Thus, for $p \geq 0$, \m{-\half \tilde M^a_a(p,p)} and 
\m{-\half \tilde M^a_a(-p,-p)} represent the quark and anti-quark 
probability densities in the baryon. Thus the baryon is a topological 
soliton in this picture. It corresponds to a static solution of the 
equations of motion (minimum of energy subject to constraints), that has 
baryon number one. A Lorentz invariant formulation is to minimize the
$(mass)^2$ of the baryon:

\vspace{.1in}
${{\cal M}^2 \over N_c^2}=\left[-\half \int p\tilde M(p,p){dp\over
2\pi}\right]\left[{-{1\over 2}} \int \tilde M(p,p){\mu^2\over
2p}{dp\over 2\pi}+{\tilde
g^2\over 8}\int dxdy |M(x,y)|^2\half|x-y|\right]$


\section{Separable or Rank One Ansatz and Valence Quark Model}

We have developed a method \cite{2dqhd} to find the minimum of energy on
the the phase space: a variant of the steepest descent method that takes 
into account the non-linear constraint. Here, we describe another method based 
on variational approximations, which brings out the connection to the
quark model. The main difficulty in minimizing the energy is the 
non-linear constraint satisfied by $M(x,y)$. We find a succession 
of variational ansatzes for $M(x,y)$ that replace this constraint with
simpler ones. These define an ascending family of sub-manifolds, which 
form a dense subset of the phase space. Minimizing the energy on 
these sub-manifolds will give us successively better approximations. 
These variational ansatzes turn out to correspond to interacting quark models.

To start with, consider an ansatz of the separable form 

\beq
\tilde  M^a_b(p,q)=-2\tilde \psi^{a}(p)\tilde \psi_b^*(q).
\eeq

\noindent This satisfies the constraint $(\eps + M)^2 =1$ if \m{\tilde \psi} 
is of norm one and of positive momentum. The Poisson Brackets of the
$M^a_b(x,y)$ imply the relations 

\beq
\{\tilde\psi_a(p),\tilde\psi_b(q)\} = 0 = 
\{\tilde\psi^{a *}(p),\tilde\psi^{b *}(q)\},
\eeq

\beq
\{\tilde \psi_a(p), \tilde\psi^{*b}(q)\} = -i2\pi\delta_a^b\delta(p-q).
\eeq

\noindent The $\tilde\psi_a$ by themselves define a classical dynamical 
system with hamiltonian 

\beq
{E_1(\psi) \over N_c} =\int_0^P\half[p+{\mu^2\over p}]
|\tilde\psi(p)|^2{dp\over 2\pi} + {\tilde g^2\over 2} 
\int |\psi(x)|^2|\psi(y)|^2{|x-y|\over 2}dxdy.
\eeq

\noindent We can quantize
this `mini' theory by looking for operators satisfying canonical 
commutation relations. Let us denote the parameter that measures the 
quantum correction, analogous to \m{\hbar},  by \m{1\over N_c}. The 
constraint on the norm can be implemented by restricting attention to 
those states $|V>$ satisfying 

\beq
\int_0^{\infty} 
\hat{\tilde\psi}^{*a}(p){\hat{\tilde\psi}}_a(p){dp\over 2\pi}|V>=1.
\eeq

\noindent A representation for our commutation relations is provided by
 bosonic creation-annihilation operators: 

\beq
[\hattilde{b}_a(p),\hattilde{b}_b(p')]=0=
[\hattilde{b}^{\dag a}(p),\hattilde{b}^{\dag b}(p')]~,~ 
[\hattilde{b}_a(p),\hattilde{b}^{\dag b}(p')]=2\pi\delta(p-q)\delta^b_a,
\eeq

\noindent with 

\beq
\hat \psi_a(x)={1\over \surd N_c} \hat b_a(x)~,~ 
\hat \psi^{\dag a}(x)=
{1\over \surd N_c} \hat b^{\dag a}(x).
\eeq

Then the constraint becomes the
condition that we restrict to states containing \m{N_c} particles:
$ \int_0^\infty \tilde b^{\dag a}(p)\tilde b_{a}(p){dp\over 2\pi}=N_c.$
Therefore, \m{N_c} must be a positive integer! Thus we are dealing 
with a system of \m{N_c} bosons interacting through a linear potential.

What are these bosons? They are the valence quarks of the parton model. 
They appear like bosons in the momentum and spin-flavour quantum numbers 
since their wave function is totally anti-symmetric in color. $N_c$ is
interpreted as the number of colors. In the {\it mean field approximation} 
\cite{ipm}, their wave function is \m{\eps^{i_1 \cdots i_{N_c}}
\tilde \psi(p_1) \cdots \tilde \psi(p_{N_c})}, which corresponds to 

\beq
|V> = a^{1 \dag}_{\tilde{\psi}} \cdots a^{N_c \dag}_{\tilde{\psi}} |0>.
\eeq

\noindent If \m{\hat \rho^a_b(p,q) = {1 \over N_c} a^{\dag ai}(p) a_{bi}(q)} 
is the quark density 
operator, then the expectation value of \m{\hat \rho(p,q)} in the 
mean field state state \m{|V>} is equal to the `classical' density matrix
\m{\tilde\rho_1(p,q) = \half (\tilde\delta(p,q) - \tilde M_1(p,q) - 
\tilde \eps(p,q))}.
Thus the classical (or large \m{N_c}) limit we have been discussing is
just the mean field approximation to this many-body problem, an idea that
goes back to Witten \cite{witten}. The semi-classical approximation will 
give us the leading corrections in the case of finite \m{N_c}. 

Since the quark null momenta are positive, their sum 
must equal the total baryon momentum \m{P}. In particular, the
parton momenta cannot exceed \m{P}. This ensures that the quark 
distributions vanish beyond \m{p = P}.
However, the total baryon momentum is extensive \m{P \sim N_c}. So in the
limit as \m{N_c\to \infty}, \m{0 \leq p<\infty}. The valence quark 
wavefunction is determined by minimizing the $(mass)^2$
subject to the normalization and momentum sum rule conditions:

\vspace{.1in}

\centerline{ \m{
{ {\cal M}^2 \over N_c^2} = \bigg[\int_0^P {p\over 2}|\tilde\psi(p)|^2
{dp\over 2\pi}\bigg]\bigg[
\int_0^P{\mu^2\over 2p}|\tilde\psi(p)|^2{dp\over 2\pi} + 
{\tilde g^2 \over 2}\int_{-\infty}^{\infty}
|\psi(x)|^2|\psi(y)|^2{|x-y| \over 2}dxdy\bigg].
}}

\vspace{.1in}

\centerline{ \m{
	\int_0^P |\tilde\psi(p)|^2{dp\over 2\pi}=1,\quad
N_c\int_0^P p|\tilde\psi(p)|^2{dp\over 2\pi}= P.
}}

\vspace{.1in}

\noindent Here $\psi(x) = \int_0^P \tilde\psi(p) e^{ipx} {dp \over {2\pi}}$.
Since we have ignored them, we will get the spin and flavor 
averaged wavefunction. The probability density of valence quarks is
\m{ V(x_B)= {P\over 2\pi} \left|\tilde\psi(x_BP)\right|^2 },
where \m{x_B = {p \over P}}.

The ground state of this many body problem was found in \cite{ipm},
in the guise of a valence quark model. Let us just summarize the result.
In the limit \m{N_c\to \infty} and \m{m=0}, the {\it absolute
minimum} of the variational principle is \m{\tilde\psi(p) = 
\sqrt{{2\pi} \over {\bar P}} e^{-{p\over{2 \bar P}}}.} Here $\bar P$
is the mean baryon momentum per color, $P \over N_c$. 
It turns out that ${\cal M} = 0$ for this configuration, so that it is 
not just a minimum on the separable submanifold, 
but on the entire phase space, in the chiral and large $N_c$ limits.
Thus we find that the ground state baryon is massless in this limit,
in agreement with Hornbostel et. al. \cite{hornbostel}.

A variational approximation to the ground state, after including the
leading effects of finite $N_c$ was also given in \cite{ipm}. In the
chiral limit, the valence quark probability distribution is 
$V(x_B)= (N_c - 1)[1-x_B]^{N_c-2}$. This variational 
approximation agrees well with our numerical solution \cite{ipm} 
and is identical to the numerical solution of Hornbostel et. al. 
(See ref. \cite{hornbostel} Eqn. 22).


\section{Rank Three Ansatz: Valence, Sea and Anti Quarks}

We can get a better approximation to the
exact soliton model, by considering a larger submanifold of the phase space,
compared to the separable ansatz, which corresponded to the valence quark 
approximation.

The departure from the valence quark picture is determined 
by the dimensionless ratio \m{{m^2\over \tilde g^2}}, a measure of
chiral symmetry breaking. Thus we should expect the anti-quark content 
to be small for small current quark masses. The leading effect of finite 
\m{N_c} is to constrain the range of momenta of the partons, as we have seen.

The mathematical advantage of the separable ansatz is that it `solves'
the  nonlinear  constraint on \m{M}: more precisely, it replaces it
with the condition that \m{\psi} is of norm one. In the same spirit,
consider the configuration 

\beq
M_r=\sum_{a,b=1}^r\xi_b^a\psi_a\otimes \psi^{\dag b}.
\eeq

\noindent Here we choose
\m{\psi_a} to be a set of \m{r} orthonormal eigenvectors of the operator
\m{\eps}; i.e., \m{\eps\psi_a=\eps_a\psi_a}, \m{\eps_a=\pm 1}. This
implies that the operator \m{M_r} is of rank \m{r}: the special case
of rank one is just the separable ansatz above. This ansatz
will satisfy the constraint on \m{M} if the \m{r\times r} matrix
\m{\xi} is hermitean and satisfies the  constraint

\beq
\xi_a^b\xi_b^c+[\eps_a+\eps_c]\xi_a^c=0,
\eeq

\noindent a `mini' version of the
constraint on \m{M}. Moreover, the baryon number is \m{B=-\half {\rm tr}\;
M= -\half tr \xi}. In the special case of rank one, we have simply
\m{\xi=-2}. By choosing a large enough value of \m{r} this ansatz can
produce as general a configuration in the phase space as needed. The 
simplest configuration of baryon number one that departs from the
separable ansatz is of rank three. We will find that for small current 
quark masses, even this departure is very small, so we do not
need to consider configurations of higher rank.

By a choice of basis among the \m{\psi_a},  we can always
bring a rank three configuration of baryon number one  to the form 

\beqs
M_3=-2\psi\otimes \psi^\dag + 
2 \zeta_-^2 [ \psi_-\otimes\psi_-^\dag -\psi_+\otimes\psi_+^\dag ] \cr
 + 2 \zeta_-\zeta_+ [\psi_-\otimes\psi_+^\dag + 
\psi_+\otimes\psi_-^\dag],
\eeqs

\noindent where \m{\psi_-,\psi,\psi_+} are three vectors in \m{L^2(R)}
 satisfying \m{\eps\psi_-=-\psi_-,\quad  \eps\psi=\psi,\quad 
\eps\psi_+=\psi_+, ||\psi_-||^2=||\psi||^2=||\psi_+||^2=1,
\quad <\psi,\psi_+>=0.}
The conditions \m{<\psi_-,\psi>=<\psi_-,\psi_+>=0} are then automatic.
The parameter \m{0\leq \zeta_- \leq 1} measures the deviation from 
the rank one ansatz and hence, the anti-quark content of the baryon. 
\m{\zeta_+ = \sqrt{1-\zeta_-^2}}.
For example, baryon number is given by 

\beq
B=\int_0^\infty\left\{|\tilde\psi(p)|^2+
\zeta_-^2\left[|\tilde\psi_+(p)|^2-|\tilde\psi_-(-p)|^2\right]\right\}
{dp\over 2\pi}.
\eeq

\noindent \m{\psi,\psi_+} vanish for $p < 0$ and 
describe valence and `sea' quarks. Their orthogonality is a consequence of
the Pauli principle. $\tilde \psi_-$ is the anti-quark wavefunction.
From our previous result we expect \m{\zeta_-} to vanish as 
\m{{m^2\over \tilde g^2}\to 0}.

This rank 3 ansatz can also be understood as arising from a
unitary transformation applied to the valence quark ansatz.
The phase space of HadronDynamics carries a transitive action of
the infinite dimensional restricted unitary group \cite{2dqhd}. 
Thus the configuration \m{M_3} can be obtained 
from \m{M_1} by a unitary transformation 

\beq
U^{\dag}(\eps - 2 \psi \otimes \psi^{\dag})U =
\eps + \xi_b^a\psi_a\otimes \psi^{\dag b}.
\eeq

\noindent Since both \m{M_1} and \m{M_3} have the same baryon number,
\m{U} lies in the connected component of the identity and
is of the form \m{U = e^{iA}} for \m{A} hermitean. From the above
expressions for \m{M_1, M_3}, we see that \m{U} is the identity except on the
span of \m{\psi_+} and \m{\psi_-}. \m{\eps = - \sigma_3}, 
in this sub-space. Thus
\m{e^{-iA} (-\sigma_3) e^{iA} = s \sigma_1 + (r-1) \sigma_3},
where \m{r = 2 \zeta_-^2} and \m{s = 2 \zeta_-\zeta_+} 
and  \m{\sigma_i} are the Pauli matrices. Therefore, on this subspace, 
\m{A} is a \m{2*2} traceless hermitean matrix 
\m{{\bf \sigma} . {\bf w}}. {\bf w} is the vector in {\bf R}$^3$ about which
\m{(0,0,-1)} must be rotated by an angle \m{2|{\bf w}|} to
reach \m{(s,0,r-1))}. Thus \m{A = {i \arcsin(\zeta_-)}
(\psi_- \otimes \psi_+^{\dag} - \psi_+ \otimes \psi_-^{\dag})}.

We can use the infinite dimensional analogue of the Pl\"ucker embedding 
\cite{Chern,Mickelsson} of the Grassmannian 
in the Fermionic Fock space to reexpress this unitary transformation on the
phase space of HadronDynamics, as a Bogoliubov transformation
on the second quantized states. The operator that corresponds to $U$
and acts on the fermionic fock space is 

\beq
\hat U = e^{i \hat A} = e^{- \arcsin(\zeta_-) (a_{i \psi_-} 
a^{i \dag}_{\psi_+} - a_{j \psi_+} a^{j \dag}_{\psi_-})},
\eeq 

\noindent the sum over colors 
produces a singlet. The angle $\theta$ of \cite{ipm} can be identified
as $\arcsin(\zeta_-)$. The second quantized
state after the Bogoliubov transformation is thus
\m{|VSA> = e^{-i \hat A} |V> }. Here $|V>$ is the
valence quark state, and $|VSA>$ stands for a state containing 
valence, sea and anti-quarks. The condition
\m{<VSA|\hat \rho(p,q)|VSA> = \tilde {\rho}_3(p,q)} is then automatic, 
since the corresponding condition was satisfied in the rank 1 case
and we have performed the same unitary transformation on both sides.

Thus, we have derived the `embellished' quark model, which contains
valence, sea and anti-quarks, as a variational approximation
to the bilocal soliton theory. The wave functions $\psi, \psi_{\pm}$
and the probability of finding an anti-quark in a baryon
${{\zeta_-^2} \over {1+ 2 \zeta_-^2}}$, were estimated in \cite{ipm}.
To summarize, we found that the probability of finding an anti-quark in 
the baryon is $ \sim 0.1 \%$ and $a \sim .035$ for 
\m{{m^2\over \tilde g^2} \sim 10^{-3}}. Moreover, the anti-quarks 
carry less than \m{0.1 \%} of the baryon momentum. 
Hornbostel et. al. \cite{hornbostel} also find a similar suppression of 
the anti-quark content in the chiral limit. 


\section{Conclusion}

Thus, while neither the quark model, nor the solitons of low energy effective 
{\it local} field theories provides a complete description of baryons
in two dimensions, a {\it bi-local} quantum field theory
provides an exact description. We have shown how
the quark model arises as a variational approximation to this 
bi-local soliton theory. Moreover, we have used this bi-local theory to
estimate the quark and anti-quark distributions in two dimensions and 
find good agreement with 
direct numerical approaches.  It is interesting to know what the 
analogous non-local theory is in four dimensions.


Acknowledgements: \DOE.


\end{document}